# Who benefits from a country's scientific research?[1]

*Giovanni Abramo (corresponding author)*

Laboratory for Studies in Research Evaluation, Institute for System Analysis and Computer Science (IASI-CNR), National Research Council of Italy
Via dei Taurini 19, 00185 Rome, Italy
giovanni.abramo@uniroma2.it

*Ciriaco Andrea D'Angelo*

Department of Engineering and Management, University of Rome "Tor Vergata" and Laboratory for Studies in Research Evaluation, Institute for System Analysis and Computer Science (IASI-CNR)
Via del Politecnico 1, 00133 Rome, Italy
dangelo@dii.uniroma2.it

**Abstract**

When a publication is cited it generates a benefit. Through the country affiliations of the citing authors, it is possible to work upwards, tracing the countries that benefit from results produced in a national research system. In this work we take the knowledge flow from Italy as an example. We develop a methodology for examination of how the knowledge flows vary across fields, in each beneficiary country. We also measure the field comparative advantage of countries in benefiting from Italian research. The results from this method can inform bilateral research collaboration policies.

# 1. Introduction

The essence of scientific activity is information processing. Scientists talk to one another, read each other's papers, and most important, they publish scientific papers. The science system consumes, transforms, produces, and exchanges "information". The aim is to produce new knowledge. Knowledge has several peculiar features compared to other goods. Knowledge is intangible, as its essence is information. It is cumulative, which means that the present global stock and level of knowledge is the direct result of scientific advancements achieved by past generations. Knowledge does not wear out physically, and can be used unlimited times without diminishing its substance: it is "infinitely expansible without loss of its intrinsic qualities, so that it can be possessed and used jointly by as many as care to do so" (David & Foray, 1995). The available stock of knowledge serves as the basis for creating new knowledge and allows for the regeneration of the existing stock, through combinations in new applications and products (Griliches, 1990). Because knowledge accumulates continuously, existing knowledge becomes obsolete and the stock must be maintained regularly.

In the current knowledge-based economy, the ability of national science systems to keep abreast and produce new scientific and technological advances is of paramount importance for sustaining domestic industrial competitiveness and socio-economic development. Access to new knowledge takes place via the channels that the scientists use to offer and disseminate it. Because the scientists' principal goal is to produce new knowledge and diffuse it, they typically encode it in publications. New knowledge spreads internationally through scientific and technical literature, seminars and conferences, and personal communication between researchers. In addition to publications, the literature recognizes social networks (Sorenson & Singh, 2007), research collaboration (Onyancha & Maluleka, 2011) and mobility of skilled persons (Kyvik & Larsen, 1997; Trippl, 2013) as important modes of knowledge transfer. The ever-growing scale and rate of dissemination beyond national boundaries stems from the ease of knowledge transmission. Owing to the rapid development of ICT, particularly the Internet, global knowledge flows have become faster, cheaper and easier than ever before.

In this work, we investigate the geographical flows of knowledge,[2] particularly from the perspective of a single country. The question we wish to answer is epitomized in the title: who benefits from a country's scientific research? While the transnational exchange of goods can be measured by the underlying monetary transactions (balance of payments), as also for the exchange of technologies ("technology balance of payments"), a problem arises when it comes to measuring knowledge flows, which do not entail commercial transactions. In this case, bibliometrics can help to overcome the problem. From the author affiliations of a publication, one can easily identify the country/countries that produced the new knowledge, and in the case of a citing publication, the country/countries that benefited from it.

To the best of our knowledge, there are very few studies on the geographic flows of "public" knowledge produced by countries. Moreover, they are very limited in scope. Rabkin, Eisemon, Lafitte-Houssat, and McLean Rathgeber (1979) explored world visibility for four departments (botany, zoology, mathematics, and physics) of the universities of Nairobi (Kenya) and Ibadan (Nigeria), measured by citations in the Science Citation Index (SCI) for the years 1963-1977. They assessed the distribution of the author-country citing publications among five macro-regions (OECD, Eastern Europe, Africa, Latin America, and Asia), with specific attention to Great Britain, given its historic relations with Kenya and Nigeria. Their findings suggested high rates of domestic visibility for scientists in the two universities, mainly in botany and zoology, which are evidently locally oriented disciplines. However, not just for these two specific disciplines, the expectation was that in general, the main recipients of new knowledge produced by a country would be domestic scholars themselves. In fact the social links between the researchers of an individual country are on average stronger than those between researchers of different countries (Bozeman & Corley, 2004), as is confirmed by observations that rates of collaboration are higher domestically than internationally (Giovanni Abramo, D'Angelo, & Murgia, 2013). At the level of the single field, Stegmann and Grohmann (2001) measured knowledge "export" and international visibility, through analysis of publication and citation data for the thirty journals listed in the Dermatology & Venereal Diseases category of the 1996 CD-ROM Journal Citation Reports (JCR) and in seven dermatology journals not listed in the 1996 JCR. Finally, Hassan and Haddawy (2013) mapped knowledge flows from the United States to other countries in the field of Energy over the years 1996-2009.

---

[2] The knowledge investigated is that encoded in publications, intentionally made available by the authors. We do not investigate flows of proprietary knowledge, such as that encoded in patents, utility models and similar, which is examined in a vast literature.

In this work, we extend the scope of previous studies, investigating the domestic and transnational flows of scientific knowledge produced in Italy, how these vary across fields, and the sectoral specialization of the countries benefiting from Italian research. The same methodology could also be applied for other countries.[3] (For the record, as of 2016, Italy ranked sixth in the world by number of publications and for number of citations.)[4]

In the next section we present the data and method of analysis. Section 3 provides the results from the elaborations. Section 4 closes the work with our considerations on the relevance of the study and its possible future developments.

## 2. Data and method

To answer the questions of who benefits from a country's scientific research, and whether differences occur across fields, we need to measure the flows of knowledge produced in the country. To this purpose, we adopt a bibliometric approach. All limitations and assumptions typical of bibliometric analyses then apply. Furthermore, from a geographical perspective, we define a publication as "made in" a source country if at least 50% of the institutions authoring it belong to that country.[5] When a publication is cited, it is conventionally understood that it has had an impact on scientific advancement because other scholars have drawn on it, more or less heavily, for the further advancement of science. We can then say that it has given rise to a "benefit". The number of "benefits" deriving from a publication equals the number of citations, and if the citing publication is co-authored by one or more foreign countries, the benefit has crossed an international boundary. In the case of a citing publication by multi-country authors, the same benefit (citation) is "gained" contemporaneously by $n$ different countries, so we can say that it has given rise to $n$ equal "gains", one for each country. Operationally, we assign a gain to each country listed in the affiliation list of the citing publication: thus, if a citing publication has three authors, two with Italy affiliations and one with France, the gains are equally assigned to both countries, independently of the number of authors in each. In theory, the total number of gains generated by a publication could be as many as the total countries in the world. A publication could be cited by $m$ publications. In this case, the publication would give rise to $m$ benefits and $m \times n$ gains. The geographical reach of a publication is measured by the total number of countries $n$ that cite it (which is lower than or equal to $m \times n$). Of course, i) the larger a country in terms of number of researchers; ii) the more productive; and iii) the more scientifically advanced in terms of domestic stock and level of accumulated knowledge, the higher the chances that it can gather benefits from new knowledge produced elsewhere.

---

[3] We plan to extend the analysis to other countries in the future. The reason why we started with Italy is that, apart from being our own country, we have Italian citing-cited publication data readily available through a license agreement with Clarivate Analytics.

[4] Latest data available from http://www.scimagojr.com/countryrank.php?year=2016&order=ci&ord=desc, last accessed 9 January 2017.

[5] It could be more correct to consider the number of authors rather than institutions, but developing appropriate algorithms would be much more complex. Alternative conventions, such as the affiliation of the corresponding author, or first and last authors in non-alphabetically ordered bylines, could be adopted as well.

In this work we analyze the geographical flows of knowledge produced in Italy and encoded in publications (articles, reviews, letters, conference proceedings) indexed in the Web of Science (WoS) over the period 2004-2008.[6] The citing publications are measured as of 31/12/2015.[7] Data for the analysis are extracted from the WoS core collection. We first download all publications published between 2004 and 2008 and authored by at least one Italian institution (271,108 in total). After excluding uncited publications and publications with less than 50% of coauthoring Italian institutions, the final dataset consists of 179,110 publications representing knowledge prevalently produced in Italy, and 2,211,772 citing (Italian and foreign) publications as of 31/12/2015.[8]

We carry out field level analysis, considering the WoS subject category (SC) identified for the journal that hosts the cited publication. We adopt a "full counting" approach, meaning that a publication published in multi-category journals is fully assigned to each SC. The cited Italian publications are distributed over 216 SCs in 13 scientific macro-areas[9] (out of a total 252 WoS SCs).

We retrieve the country names of citing publications based on the affiliations of the authors. The resulting list includes two countries that have since subdivided: Yugoslavia and Serbia-Montenegro. We observe the dates of these events and reassign the subsequent scientific production among the current countries.[10] The People's Republic of China (mainland China), Hong Kong and Macau are merged as China; England, Wales, Scotland, Northern Ireland and Gibraltar are merged as the United Kingdom. Citations from France, the Netherlands and New Zealand are merged with those of their overseas territories. Addresses of citing publications without country indication (seven in all) were excluded. The final number of citing countries is 197.

---

[6] The breadth of the period of observation (five years) ensures sufficient robustness of results (Abramo, D'Angelo, & Cicero, 2012).

[7] The breadth of the citation window (seven years from last date of publication) ensures robust results (Abramo, Cicero, & D'Angelo, 2011).

[8] Citing publications are not limited to any particular type of document.

[9] Mathematics; Physics; Chemistry; Earth and Space Sciences; Biology; Biomedical Research; Clinical Medicine; Psychology; Engineering; Economics; Law, political and social sciences; Art and Humanities; Multidisciplinary Sciences. The macro-areas and the assignment of SCs to them were at some point defined by ISI (now Clarivate), although no longer showing in Clarivate bilbiometric products. There is no multi-assignment of SCs to macro-areas.

## 3. Results and analysis

The 179,110 2004-2008 Italian publications were cited by 2,211,772 publications. Since each citing paper cites on average 1.66 Italian publications, total benefits (to the end of 2015) amounted to 3,666,633. Each benefit was earned on average by 1.4 countries, so the total amount of gains was 5,124,147. In the following Subsection, we present the distribution of gains among countries, at both the aggregate and SC level. In Subsection 3.2, we analyze the geographical reach of Italian scientific research, i.e. the number of countries that benefited from the results of the research, once again at both aggregate and SC level. In Subsection 3.3, we analyze the comparative advantage of countries at benefiting from Italian research, through measurement of a specialization index.

### 3.1 Distribution of gains among countries

For each citing country, we calculate the total number of gains. Table 1 shows the top 50 beneficiary countries by number of gains. The rank by number of gains is compared to the world rank by total number of WoS publications in the 2004-2015 period.[11] Italy holds the lion's share of gains (19.4%), notwithstanding it ranks eighth for total number of publications produced. USA follows with 16.5%, then China (6.6%), Germany (5.9%) and UK (5.6%). Among the top 50 beneficiary countries, in six cases the gains and publication rank are aligned; the maximum rank shift is observed for Italy (+7 positions) and Russia (-7 positions). The correlation coefficient between the two ranks is very high (Spearman ρ between ranks of column 4 and 7 is 0.970), which means that the accumulated gain of a country is unmistakably correlated to its "scientific size". The positive shift of Italy can be possibly ascribed to the following factors: i) self-citations; ii) geographical and social proximity; iii) research oriented towards domestic needs. Geographical and social proximity may explain the positive shift of Greece. The negative shift of Russia reveals instead that Italy plays a secondary role in their citation networks, or they specialize in research fields with lower citation intensity, or both.

---

[10] What remained of the Federal Republic of Yugoslavia was officially renamed Serbia-Montenegro in 2003; citing publications from Yugoslavia or Serbia-Montenegro were therefore summed. In 2006 Serbia-Montenegro broke up: the summed citations were thus reassigned to the new countries on the basis of the relative shares of benefits accumulated to these countries by the end of period under examination.

[11] Numbers of WoS publications were extracted from InCites™, a customized, citation-based research analytics tool made available by Clarivate Analytics.

*Table 1: Top 50 countries for gains generated by 2004-2008 WoS Italian publications (citations observed 31/12/2015).*

| Beneficiary country | Gains | Ratio to total gains (%) | Rank | 2004-2015 WoS publications | Ratio to total publications (%) | Rank | Gains / tot. publications | Rank |
|---|---|---|---|---|---|---|---|---|
| Italy | 996635 | 19.4% | 1 | 948437 | 3.3% | 8 | 1.051 | 1 |
| USA | 844953 | 16.5% | 2 | 6964383 | 24.6% | 1 | 0.121 | 43 |
| China | 336669 | 6.6% | 3 | 2555757 | 9.0% | 2 | 0.132 | 38 |
| Germany | 301071 | 5.9% | 4 | 1590219 | 5.6% | 4 | 0.189 | 18 |
| United Kingdom | 285329 | 5.6% | 5 | 1906350 | 6.7% | 3 | 0.150 | 33 |
| France | 239572 | 4.7% | 6 | 1081642 | 3.8% | 6 | 0.221 | 9 |
| Spain | 201725 | 3.9% | 7 | 762209 | 2.7% | 9 | 0.265 | 3 |
| Japan | 153037 | 3.0% | 8 | 1329594 | 4.7% | 5 | 0.115 | 47 |
| Canada | 140268 | 2.7% | 9 | 1004387 | 3.5% | 7 | 0.140 | 34 |
| Netherlands | 104909 | 2.0% | 10 | 550905 | 1.9% | 13 | 0.190 | 16 |
| Australia | 103041 | 2.0% | 11 | 742274 | 2.6% | 10 | 0.139 | 36 |
| Switzerland | 90641 | 1.8% | 12 | 390935 | 1.4% | 16 | 0.232 | 7 |
| India | 87190 | 1.7% | 13 | 679195 | 2.4% | 11 | 0.128 | 41 |
| Brazil | 81017 | 1.6% | 14 | 482393 | 1.7% | 14 | 0.168 | 27 |
| South Korea | 76988 | 1.5% | 15 | 639641 | 2.3% | 12 | 0.120 | 44 |
| Belgium | 65118 | 1.3% | 16 | 299453 | 1.1% | 21 | 0.217 | 12 |
| Poland | 61638 | 1.2% | 17 | 324326 | 1.1% | 20 | 0.190 | 17 |
| Sweden | 59252 | 1.2% | 18 | 344454 | 1.2% | 19 | 0.172 | 25 |
| Turkey | 55246 | 1.1% | 19 | 344545 | 1.2% | 18 | 0.160 | 30 |
| Greece | 49363 | 1.0% | 20 | 184025 | 0.6% | 26 | 0.268 | 2 |
| Taiwan | 48868 | 1.0% | 21 | 379068 | 1.3% | 17 | 0.129 | 40 |
| Russia | 47264 | 0.9% | 22 | 424202 | 1.5% | 15 | 0.111 | 48 |
| Austria | 47223 | 0.9% | 23 | 214937 | 0.8% | 23 | 0.220 | 11 |
| Portugal | 40842 | 0.8% | 24 | 173776 | 0.6% | 28 | 0.235 | 5 |
| Iran | 40266 | 0.8% | 25 | 253116 | 0.9% | 22 | 0.159 | 31 |
| Denmark | 40177 | 0.8% | 26 | 212716 | 0.8% | 24 | 0.189 | 19 |
| Israel | 37585 | 0.7% | 27 | 206069 | 0.7% | 25 | 0.182 | 20 |
| Finland | 30564 | 0.6% | 28 | 172116 | 0.6% | 29 | 0.178 | 21 |
| Czech Republic | 29191 | 0.6% | 29 | 175001 | 0.6% | 27 | 0.167 | 28 |
| Norway | 26827 | 0.5% | 30 | 159538 | 0.6% | 32 | 0.168 | 26 |
| Mexico | 26293 | 0.5% | 31 | 162167 | 0.6% | 30 | 0.162 | 29 |
| Argentina | 22218 | 0.4% | 32 | 111229 | 0.4% | 38 | 0.200 | 15 |
| Ireland | 21733 | 0.4% | 33 | 125163 | 0.4% | 35 | 0.174 | 24 |
| Hungary | 21121 | 0.4% | 34 | 99801 | 0.4% | 39 | 0.212 | 13 |
| Singapore | 20619 | 0.4% | 35 | 159720 | 0.6% | 31 | 0.129 | 39 |
| Chile | 18518 | 0.4% | 36 | 77402 | 0.3% | 42 | 0.239 | 4 |
| Romania | 17816 | 0.3% | 37 | 128219 | 0.5% | 34 | 0.139 | 35 |
| New Zealand | 16415 | 0.3% | 38 | 123495 | 0.4% | 36 | 0.133 | 37 |
| South Africa | 16115 | 0.3% | 39 | 135222 | 0.5% | 33 | 0.119 | 45 |
| Egypt | 15695 | 0.3% | 40 | 88818 | 0.3% | 40 | 0.177 | 22 |
| Saudi Arabia | 13655 | 0.3% | 41 | 77388 | 0.3% | 43 | 0.176 | 23 |
| Malaysia | 12669 | 0.2% | 42 | 118490 | 0.4% | 37 | 0.107 | 50 |
| Serbia | 12082 | 0.2% | 43 | 54850 | 0.2% | 46 | 0.220 | 10 |
| Slovenia | 11694 | 0.2% | 44 | 50065 | 0.2% | 49 | 0.234 | 6 |
| Croatia | 11465 | 0.2% | 45 | 50729 | 0.2% | 48 | 0.226 | 8 |
| Thailand | 10300 | 0.2% | 46 | 88598 | 0.3% | 41 | 0.116 | 46 |
| Tunisia | 8765 | 0.2% | 47 | 42385 | 0.1% | 50 | 0.207 | 14 |

| | | | | | | | |
|---|---|---|---|---|---|---|---|
| Slovakia | 8370 | 0.2% | 48 | 54520 | 0.2% | 47 | 0.154 | 32 |
| Pakistan | 8186 | 0.2% | 49 | 64540 | 0.2% | 45 | 0.127 | 42 |
| Ukraine | 7944 | 0.2% | 50 | 74263 | 0.3% | 44 | 0.107 | 49 |

We now turn to field level analysis. To assess possible differences across fields, we repeat the same analysis in each SC. As an example, in Figure 1 we show the geographic distribution of gains generated by Italian scientific research in Tropical Medicine. A total of 118 countries gained from Italian research. Gains are distributed unevenly, with the highest shares appropriated by Italy (13.8%) and USA (11.4%).

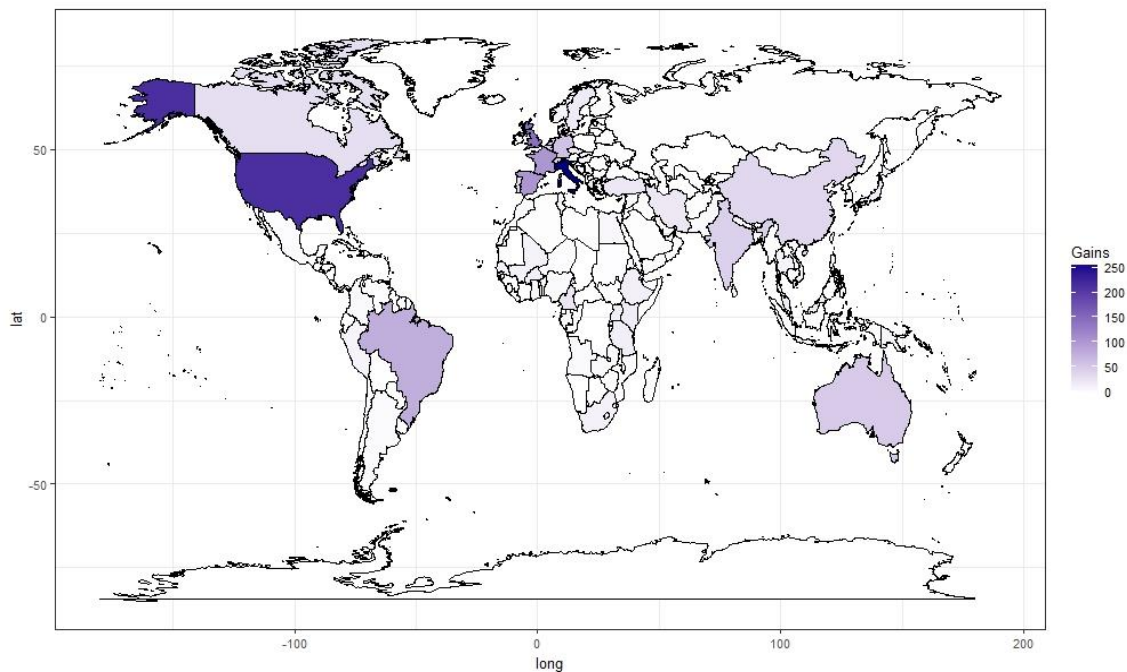

*Figure 1: Geographic distribution of gains generated by Italian 2004-2008 WoS publications in Tropical Medicine*

Table 2 presents a summary of statistics for all 216 SCs, listing all nations ranking among the top five by number of gains in at least an SC. Only 23 countries (12% of total) reached "top five" in number of gains for at least one SC, and of these only half reached this status in five or more SCs. Italy and the USA dominate: Italy as the largest recipient of gains in 149 SCs (69.0%), and USA in 62 (28.7%). Only two other countries rank at the very top in at least one SC: the UK in three SCs, all belonging to the social sciences (International relation; Law; Public administration), and China in two (Engineering, industrial; Metallurgy & metallurgical engineering). The latter is not surprising, given the rapid industrial growth of China in recent years.

In SCs where Italy is not dominant, it is either second (in 63 SCs) or third (in 4) in share of gains; the USA follows closely, showing only one SC (Architecture & Art) where it falls below fifth position.

*Table 2: List of countries ranking among top five by number of gains in at least one subject category, and rank frequencies*

| Country | 1-st | 2-nd | 3-rd | 4-th | 5-th | Total |
|---|---|---|---|---|---|---|
| Italy | 149 | 63 | 4 | 0 | 0 | 216 |
| USA | 62 | 105 | 47 | 1 | 0 | 215 |
| United Kingdom | 3 | 8 | 79 | 41 | 31 | 162 |
| Germany | 0 | 0 | 22 | 85 | 36 | 143 |
| China | 2 | 36 | 44 | 15 | 22 | 119 |
| France | 0 | 0 | 5 | 23 | 67 | 95 |
| Spain | 0 | 4 | 11 | 32 | 21 | 68 |
| Canada | 0 | 0 | 1 | 10 | 10 | 21 |
| Netherlands | 0 | 0 | 0 | 4 | 12 | 16 |
| India | 0 | 0 | 1 | 3 | 5 | 9 |
| Australia | 0 | 0 | 0 | 1 | 4 | 5 |
| Brazil | 0 | 0 | 1 | 0 | 1 | 2 |
| Turkey | 0 | 0 | 1 | 0 | 1 | 2 |
| Switzerland | 0 | 0 | 1 | 0 | 0 | 1 |
| Estonia | 0 | 0 | 0 | 1 | 0 | 1 |
| Portugal | 0 | 0 | 0 | 1 | 0 | 1 |
| Belgium | 0 | 0 | 0 | 0 | 1 | 1 |
| Greece | 0 | 0 | 0 | 0 | 1 | 1 |
| Iran | 0 | 0 | 0 | 0 | 1 | 1 |
| Israel | 0 | 0 | 0 | 0 | 1 | 1 |
| Japan | 0 | 0 | 0 | 0 | 1 | 1 |
| South Korea | 0 | 0 | 0 | 0 | 1 | 1 |
| Sweden | 0 | 0 | 0 | 0 | 1 | 1 |

Next, we explore the complementary aspect of gains earned by Italy and USA, per SC. Tables 3 and 4 respectively show the SCs with the highest and the lowest ratio of national gains out of total benefits, for Italy and USA. From the second row of Table 3, we learn that 791 prevalently Italian publications in Geology were published between 2004 and 2008. Further, 6330 publications cited them (benefits). Of this total, 5701 of the citing publications were authored by Italian institutions (gains to Italy), thus Italy appropriated 90.1% of benefits embedded in its publications in Geology. The highest gains to benefits ratio occur in the SCs of Earth sciences, which it is mainly a locally-oriented research area, while USA benefited most in the SCs belonging to Clinical Medicine and Biomedical Research. The lowest gains by Italy were in the macro-area of Law, political and social sciences; for the USA this occurred in Art and Humanities and in Materials Science categories. It must be noted that in those SCs where the skewness of gains for Italy is lower, Italy still remains among the main beneficiaries: in Medieval & Renaissance Studies Italy ranks top by total gains; while in the other nine bottom-ranked SCs it places 2nd or 3rd. The countries which obtain high number of gains in these SCs are the ones listed in Table 2. The lower share of gains obtained by Italy is likely due to the fact that, differently from the Sciences, Italian scientists in these SCs publish both in national and international journals, and definitely less than English speaking scientists in international journals. In these SCs it is more likely then that citing publications are authored by foreign (English-speaking) countries.

*Table 3: Subject categories (in Italy) with the highest and the lowest gains to benefits ratio*

| Subject category | Macro-area* | Italian publications | Total benefits (a) | Total Italian gains (b) | Ratio % (b/a) |
|---|---|---|---|---|---|
| Geology | 4 | 791 | 6330 | 5701 | 90.1% |
| Geochemistry & Geophysics | 4 | 2296 | 26156 | 22213 | 84.9% |
| Geosciences, Multidisciplinary | 4 | 2986 | 35053 | 28231 | 80.5% |
| Nuclear Science & Technology | 9 | 2319 | 14847 | 11066 | 74.5% |
| Paleontology | 4 | 435 | 4028 | 2950 | 73.2% |
| Mineralogy | 4 | 662 | 6465 | 4718 | 73.0% |
| Astronomy & Astrophysics | 2 | 4755 | 60502 | 43199 | 71.4% |
| Mathematics | 1 | 3516 | 20828 | 13048 | 62.6% |
| Physics, Nuclear | 2 | 1430 | 14435 | 8728 | 60.5% |
| Architecture & Art | 12 | 79 | 260 | 157 | 60.4% |
| --- | | | | | |
| Political Science | 11 | 367 | 3189 | 622 | 19.5% |
| Integrative & Complementary Medicine | 7 | 145 | 2935 | 547 | 18.6% |
| Multidisciplinary Sciences | 13 | 104 | 8769 | 1544 | 17.6% |
| Religion | 12 | 64 | 142 | 25 | 17.6% |
| Medieval & Renaissance Studies | 12 | 26 | 52 | 9 | 17.3% |
| Law | 11 | 199 | 978 | 169 | 17.3% |
| Literature | 12 | 96 | 200 | 31 | 15.5% |
| Sociology | 11 | 149 | 1778 | 266 | 15.0% |
| Classics | 12 | 75 | 136 | 20 | 14.7% |
| International Relations | 11 | 120 | 971 | 122 | 12.6% |

*\* 1, Mathematics; 2, Physics; 3, Chemistry; 4, Earth and Space Sciences; 5, Biology; 6, Biomedical Research; 7, Clinical Medicine; 8, Psychology; 9, Engineering; 10, Economics; 11, Law, political and social sciences; 12, Art and Humanities; 13, Multidisciplinary Sciences.*

*Table 4: Subject categories (in USA) with the highest and the lowest gains to benefits ratio*

| Subject category | Macro-area* | Italian publications | Total benefits (a) | Total US gains (b) | Ratio % (b/a) |
|---|---|---|---|---|---|
| Astronomy & Astrophysics | 2 | 4755 | 60502 | 43598 | 72.1% |
| Substance Abuse | 7 | 144 | 2632 | 1257 | 47.8% |
| Neurosciences | 7 | 6740 | 134094 | 58681 | 43.8% |
| Oncology | 6 | 7642 | 153770 | 65744 | 42.8% |
| Psychology, Psychoanalysis | 8 | 62 | 397 | 169 | 42.6% |
| Psychiatry | 7 | 1711 | 32173 | 13628 | 42.4% |
| Hematology | 6 | 3732 | 80587 | 33916 | 42.1% |
| Cardiac & Cardiovascular Systems | 7 | 4289 | 79939 | 32989 | 41.3% |
| Cell Biology | 5 | 4467 | 132194 | 53899 | 40.8% |
| Immunology | 6 | 4571 | 92406 | 37466 | 40.5% |
| --- | | | | | |
| Materials Science, Coatings & Films | 9 | 774 | 9408 | 1110 | 11.8% |
| Food Science & Technology | 5 | 2815 | 45620 | 5325 | 11.7% |
| Logic | 1 | 167 | 1172 | 134 | 11.4% |
| Metallurgy & Metallurgical Engineering | 9 | 720 | 8056 | 921 | 11.4% |
| Materials Science, Composites | 9 | 373 | 5160 | 586 | 11.4% |
| Agriculture, Multidisciplinary | 5 | 1250 | 17505 | 1883 | 10.8% |
| Chemistry, Applied | 3 | 1906 | 37502 | 3922 | 10.5% |
| Art | 12 | 204 | 1271 | 106 | 8.3% |
| Materials Science, Textiles, Paper & Wood | 9 | 39 | 427 | 30 | 7.0% |
| Architecture & Art | 12 | 79 | 260 | 8 | 3.1% |

*\* Same as in Table 3.*

## 3.2 Geographical reach of Italian scientific research

As of the close of 2015, the 2004-2008 Italian scientific production had been cited by authors affiliated with institutions of 197 countries, out of the 204 indexed by InCites. In this section we present the results of the analysis of the geographical reach of (i.e. the number of countries that benefited from) Italian research in each SC. We expect that the higher the number and impact of cited publications in an SC, the higher will be the geographical reach of the SC. In fact the rank correlation coefficient (Spearman ρ) between geographical reach and number of publications is 0.812, and between geographical reach and average impact (average benefit per publication of the SC) the coefficient is 0.514.

In Table 5 we show the top ten and bottom ten SCs for geographical reach. Biochemistry & Molecular Biology has the highest value: the 8554 Italian publications in this subject category are cited in publications by institutions from a full 174 different countries. Following this are Environmental Sciences (167), then three SCs in Biomedical Research and two in Biology, which show geographical reach between 160 and 165 countries. On the opposite front, at the very bottom, we find Medieval and Renaissance Studies, where the 26 Italian works have been cited by authors in only 10 countries. In fact the lower part of the table is dominated by SCs in the Arts and Humanities, consistently with less than 100 publications cited, and never exceeding a geographical reach of 41 nations.

*Table 5: Subject categories with the highest and the lowest geographical reach of Italian scientific research*

| Subject category | Macro-area* | Italian publications | Average impact (benefit per publication) | Geographical reach |
|---|---|---|---|---|
| Biochemistry & Molecular Biology | 5 | 8554 | 23.6 | 174 |
| Environmental Sciences | 4 | 3854 | 15.3 | 167 |
| Infectious Diseases | 6 | 1934 | 15.5 | 165 |
| Pharmacology & Pharmacy | 6 | 6596 | 18.9 | 165 |
| Immunology | 6 | 4571 | 20.2 | 163 |
| Microbiology | 5 | 2504 | 18.9 | 163 |
| Ecology | 5 | 1307 | 18.3 | 160 |
| Geosciences, Multidisciplinary | 4 | 2986 | 11.7 | 158 |
| Genetics & Heredity | 7 | 2722 | 20.8 | 157 |
| Endocrinology & Metabolism | 7 | 4024 | 21.2 | 156 |
| … | | | | |
| Engineering, Marine | 9 | 56 | 5.7 | 41 |
| Architecture & Art | 12 | 79 | 3.3 | 38 |
| Literature | 12 | 96 | 2.1 | 38 |
| Dance, Theater, Music, Film, Folklore | 12 | 55 | 4.0 | 34 |
| Psychology, Psychoanalysis | 8 | 62 | 6.4 | 34 |
| History of Social Sciences | 11 | 62 | 4.7 | 32 |
| Humanities, Multidisciplinary | 12 | 45 | 3.1 | 25 |
| Religion | 12 | 64 | 2.2 | 20 |
| Classics | 12 | 75 | 1.8 | 19 |
| Medieval & Renaissance Studies | 12 | 26 | 2.0 | 10 |

* *1, Mathematics; 2, Physics; 3, Chemistry; 4, Earth and Space Sciences; 5, Biology; 6, Biomedical Research; 7, Clinical Medicine; 8, Psychology; 9, Engineering; 10, Economics; 11, Law, political and social sciences; 12, Art and Humanities; 13, Multidisciplinary Sciences.*

Figure 2 shows the range of variation in geographical reach for the SCs in each disciplinary macro-area, along with the overall average value (104, the horizontal line).

The lowest number of citing countries (10) is observed in Art and Humanities, while the maximum (174) is in Biology. The lowest variability within the macro-area is observed for Chemistry (107-143), and the highest is in Biology (76-174).

To control for the number of cited publications, we extract from each SC 100 random samples, each consisting of 100 cited publications. For each sample, we measure the geographical reach. We then average the 100 values of the geographical reach in each SC. Figure 3 shows the range of variation of the average geographical reach observed in such samples within each macro-area. The smallest range (64.8-72.7) is observed in Mathematics (58-83.6) and the largest in Art and Humanities (10-67.3) Clinical Medicine (62.2-128.6) and Law, political and social sciences (32-97.6). The average value across macro-areas is 80.5 (horizontal line).

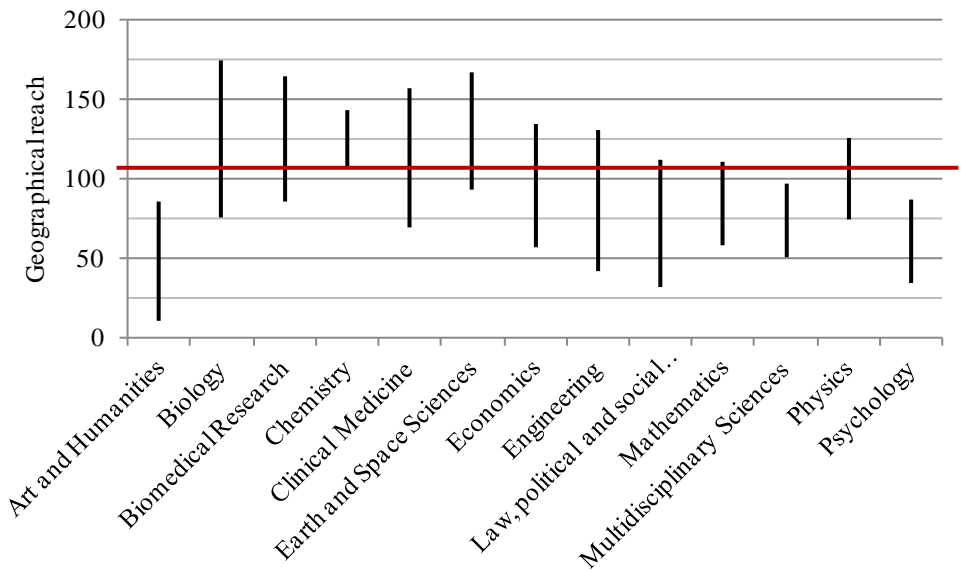

*Figure 2: Range of variation in geographical reach for the disciplinary macro-areas*

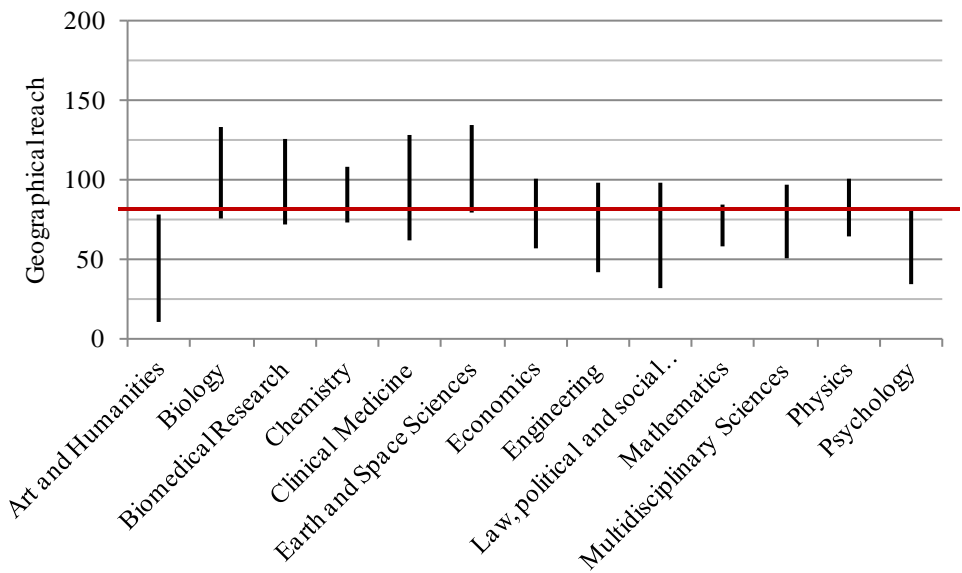

*Figure 3: Range of variation of geographical reach for 100 random samples of 100 cited publications in each SCs, per macro-area*

### 3.3 The comparative advantage of countries at benefiting from Italian research

In this subsection, we answer the last research question on the capability of countries to appropriate benefits across Italian research SCs, as compared to the rest of the world. We measure this sectoral specialization through an indicator named Scientific Gain Specialization Index (SGSI). SGSI measures a country's capacity to benefit from another country's research as compared to the rest of the world, across all research fields. SGSI is conceptually similar to the merchandise trade specialization index, whereby merchandise is replaced by knowledge. In operational terms, SGSI is calculated applying the Revealed Comparative Advantage (RCA) methodology and, in particular, the *Balassa index* (Balassa, 1979). The SGSI of country k in the SCj (SGSI$_{kj}$) is defined as:

$$SGSI_{kj} = 100 * \tanh ln \left\{ \frac{(G_{kj}/\sum_j G_{kj})}{\sum_k G_{kj} / \sum_k \sum_j G_{kj}} \right\}$$

with $G_{kj}$ indicating the gains obtained by country *k* in the SC*j*. Use of the logarithmic function centers the data around zero and the hyperbolic tangent multiplied by 100 limits the $SSI_{kj}$ values to a range of +100 to -100 (Giovanni Abramo, D'Angelo, & Di Costa, 2014). For any SC, the closer the value of the index to +100 the more the country is specialized in that SC in capturing benefits from new knowledge produced by another country (or internally). Vice versa, the closer the index approaches -100, the less the country is specialized in the SC. Values around 0 are labeled as "expected".

We adopt two perspectives: one domestic and one international. For the Italian case, Table 6 shows the top and bottom ten SCs by SGSI value, together with the data for index calculation. Italy's specialization in benefiting from its own research, vis-à-vis the rest of the world, is in Architecture & art and in four SCs with strongly domestic areas of application (Geology; Engineering, marine; Geochemistry & geophysics; Geosciences, multidisciplinary). Overall, there are 24 SCs (11.1% of total, 216) with values of SGSI above +30 and ten SCs (4.6%) showing SGSI below -30. The lowest value of SGSI (-56.4) occurs in International relations; two further SCs (Management and Business, all three in Economics macro-area) show SGSI values below -35. These findings align with those reported in Table 2, with particular reference to the UK and China.

Table 7 instead shows the top and bottom ten SCs by SGSI for the US. In this case we note the presence of the Psychology macro-area in the upper part of the SCs ranking, indicating strong interest from American scientists concerning the research conducted in this field by their Italian colleagues.

This analysis can be replicated for each country, to identify the SCs in which each country shows the highest comparative advantages in gaining from Italian research. To exemplify, Brazil shows the highest SGSI (94.1) in Dentistry, oral surgery & medicine; India in Integrative & complementary medicine (84.3); the Netherlands in Medieval & renaissance studies (93.0); South Korea in Engineering, marine (80.9) and so on.

*Table 6: Ten top and bottom-ranked SC by scientific gain specialization index (SGSI), for gains within Italy*

| Subject category | Total cited publications | Gains for Italy | Total gains | Specialization index (SGSI) |
|---|---|---|---|---|
| Architecture & Art | 79 | 157 | 332 | 70.2 |
| Geology | 791 | 5701 | 14318 | 60.4 |
| History of Social Sciences | 62 | 118 | 342 | 50.5 |
| Engineering, Marine | 56 | 164 | 482 | 49.4 |
| Geochemistry & Geophysics | 2296 | 22213 | 66213 | 48.4 |
| Art | 204 | 618 | 1870 | 47.2 |
| Geosciences, Multidisciplinary | 2986 | 28231 | 86187 | 46.5 |
| Mathematics | 3516 | 13048 | 40374 | 45.5 |
| Materials Science, Characterization & Testing | 360 | 913 | 2827 | 45.4 |
| Logic | 167 | 535 | 1658 | 45.3 |
| --- | | | | |
| Respiratory System | 1636 | 7606 | 53036 | -31.1 |
| Engineering, Industrial | 611 | 1718 | 12090 | -32.0 |
| Tropical Medicine | 63 | 259 | 1878 | -34.6 |
| Management | 603 | 1992 | 14718 | -36.3 |
| Business | 315 | 1035 | 7676 | -36.6 |
| Ophthalmology | 941 | 2629 | 19834 | -38.1 |
| Multidisciplinary Sciences | 104 | 1544 | 12188 | -41.9 |
| Sociology | 149 | 266 | 2180 | -44.9 |
| Integrative & Complementary Medicine | 145 | 547 | 4562 | -46.3 |
| International Relations | 120 | 122 | 1167 | -56.4 |

*Table 7: Top and bottom ten ranked SCs by SGSI, for gains to the USA*

| Subject category | Total cited publications | Gains for Italy | Total gains | Specialization index (SGSI) |
|---|---|---|---|---|
| Psychology, Psychoanalysis | 62 | 169 | 486 | 64.9 |
| Substance Abuse | 144 | 1257 | 4120 | 56.7 |
| Religion | 64 | 33 | 120 | 49.3 |
| Multidisciplinary Sciences | 104 | 3184 | 12188 | 45.3 |
| Psychology, Clinical | 232 | 1653 | 6529 | 42.8 |
| Psychology, Social | 166 | 900 | 3596 | 41.8 |
| Cell Biology | 4467 | 53899 | 215453 | 41.8 |
| Psychology, Developmental | 185 | 1399 | 5596 | 41.7 |
| Humanities, Multidisciplinary | 45 | 31 | 125 | 41.1 |
| Psychiatry | 1711 | 13628 | 55269 | 40.6 |
| --- | | | | |
| Agriculture, Dairy & Animal Science | 1105 | 1192 | 15317 | -61.9 |
| Archaeology | 352 | 372 | 4823 | -62.4 |
| Nuclear Science & Technology | 2319 | 2812 | 36470 | -62.4 |
| Fisheries | 337 | 638 | 8305 | -62.7 |
| Agriculture, Multidisciplinary | 1250 | 1883 | 26591 | -67.3 |
| Chemistry, Applied | 1906 | 3922 | 56956 | -68.8 |
| Food Science & Technology | 2815 | 5325 | 78588 | -69.7 |
| Art | 204 | 106 | 1870 | -77.8 |
| Materials Science, Textiles, Paper & Wood | 39 | 30 | 560 | -79.9 |
| Architecture & Art | 79 | 8 | 332 | -95.6 |

## 4. Conclusions

Most studies on international knowledge flows focus on cross-sector flows, particularly public-to-private knowledge flows, within a wider technology transfer perspective (Reddy & Zhao, 1990). Another current of studies in flows of knowledge enters under the umbrella of literature on international R&D collaboration (Bozeman, Fay, & Slade, 2013). Very few works, all limited in scope, concern the vertical international knowledge flows within the scientific community. We aim to start filling the void with this initial investigation, which develops a methodology and begins from the Italian example. The extension of the approach to other nations would be straightforward.

Starting from the national scientific output indexed in WoS between 2004 and 2008, we have traced all citing publications upward, thus identifying all citing countries who benefited from Italian research over the period observed. There were 197 such countries to the end of 2015, out of the 204 indexed by InCites. As expected, we found a high correlation between the research size of recipient countries and their ability to benefit from Italian research. Also as expected, Italy results as the main beneficiary of its own research results, explained by concomitant factors: size and level of domestic stock of knowledge, self-citations, higher intensity of domestic collaborations favored by social and geographical proximity, and orientation of some research activities towards knowledge uniquely relevant to the given national context.

Next, through a Scientific Gain Specialization Index, we were able to measure the comparative advantage of single countries in benefiting from Italian research, field by field. In comparing between fields, this analysis reveals the ones in which Italy (or another country under observation) benefits from its own research more than do other countries.

The methodology developed provides useful results for informing national research strategies, for example the analysis of comparative advantage of foreign countries could be particularly pertinent concerning bilateral collaboration. Extending the observation period, would allow cross-time analysis to monitor how such comparative advantages vary along time. Thinking of a very few cases, such analyses could be of interest to diplomatic attachés, or trade negotiators dealing with scientific issues.

In future research, we intend to extend the analysis to two or more other countries, then being able to carry out comparisons between countries in terms of their incoming and outgoing knowledge flows. The ultimate goal would be to measure the balance of knowledge flows for all countries, paralleling the measurement of technology balance of payments, and make this alongside other yearly reports of science and technology indicators.